\documentclass[journal]{IEEEtran}

\usepackage{pgfplots}
\usepackage[bookmarks,colorlinks]{hyperref}
\usepackage[linesnumbered,ruled,lined]{algorithm2e}
\usetikzlibrary{shapes.multipart,intersections}
\usepackage{amsmath,amssymb,amsfonts,amsthm,steinmetz}
\usepackage{mathrsfs}  
\usepackage{textcomp}
\usepackage{acronym}
\usepackage{xcolor}
\usepackage{upgreek,xspace}
\usepackage{array}
\usepackage{tikz}
\usetikzlibrary{calc}
\makeatletter
\newcommand{\gettikzxy}[3]{%
  \tikz@scan@one@point\pgfutil@firstofone#1\relax
  \edef#2{\the\pgf@x}%
  \edef#3{\the\pgf@y}%
}
\makeatother

\usepackage{comment}

\usepackage[draft]{todonotes}   
\usepackage[T1]{fontenc}

\usepackage{esvect}

\usepackage{times}
\usepackage{bm}
\usepackage{stmaryrd}
\usepackage{babel}
\usepackage{graphics, graphicx}
\usepackage{gensymb}
\usepackage{cite}
\usepackage{enumitem}
\usepackage{url}

\begin{document}

\title{Electromagnetic Bounds on Realizing Targeted MIMO Transfer Functions in Real-World  Systems with Wave-Domain Programmability}

\author{Philipp~del~Hougne,~\IEEEmembership{Member,~IEEE}
\thanks{
P.~del~Hougne is with the Department of Electronics and Nanoengineering, Aalto University, 00076 Espoo, Finland and with Univ Rennes, CNRS, IETR - UMR 6164, F-35000, Rennes, France. (e-mail:philipp.del-hougne@univ-rennes.fr)
}
\thanks{This work was supported in part by the Nokia Foundation (project 20260028), the ANR France 2030 program (project ANR-22-PEFT-0005), the ANR PRCI program (project ANR-22-CE93-0010), the Rennes M\'etropole AES program (project ``SRI''), the European Union's European Regional Development Fund, and the French region of Brittany and Rennes Métropole through the contrats de plan État-Région program (projects ``SOPHIE/STIC \& Ondes'' and ``CyMoCoD'').}
}

\maketitle

\begin{abstract}
A key question for most applications involving reconfigurable linear wave systems is how accurately a desired linear operator can be realized by configuring the system's tunable elements. The relevance of this question spans from hybrid-MIMO analog combiners via computational meta-imagers to programmable wave-domain signal processing. Yet, no electromagnetically consistent bounds have been derived for the fidelity with which a desired operator can be realized in a real-world reconfigurable wave system. Here, we derive such bounds based on an electromagnetically consistent multiport-network model (capturing mutual coupling between tunable elements) and accounting for real-world hardware constraints (lossy, 1-bit-programmable elements). Specifically, we formulate the operator-synthesis task as a quadratically constrained fractional-quadratic problem and compute rigorous fidelity upper bounds based on semidefinite relaxation. 
We apply our technique to three distinct experimental setups. The first two setups are, respectively, a free-space and a rich-scattering $4\times 4$ MIMO channel at 2.45~GHz parameterized by a reconfigurable intelligent surface (RIS) comprising 100 1-bit-programmable elements. The third setup is a $4\times 4$ MIMO channel at 19~GHz from four feeds of a dynamic metasurface antenna (DMA) to four users. 
We systematically study how the achievable fidelity scales with the number of tunable elements, and we probe the tightness of our bounds by trying to find optimized configurations approaching the bounds with standard discrete-optimization techniques. 
We observe a strong influence of the coupling strength between tunable elements on our fidelity bound. For the two RIS-based setups, our bound attests to insufficient wave-domain flexibility for the considered operator synthesis.
\end{abstract}


\section{Introduction}
\label{sec_introduction}

Linear wave systems comprising tunable elements in the wave domain play an increasingly central role in applications spanning the areas of wireless communications, sensing, and computing~\cite{pWDCperspective}. These reconfigurable linear wave systems realize a programmable linear operation on the input wavefront. However, the extent to which a given real-world reconfigurable linear wave system is capable of realizing a specific desired linear transformation is generally unknown. Optimizing the tunable elements' configuration is typically a discrete, high-dimensional, and non-convex problem with unknown global solution. In a typical scenario with 100 1-bit-tunable elements, there are $2^{100}$ possible configurations. Consequently, it is difficult to assess the outcome of a chosen optimization technique, and it remains generally unclear whether there could be other optimization techniques reaching substantially better solutions. Optimizing the configuration toward the desired functionality thus yields some heuristic insights but important fundamental conclusions remain elusive.

However, knowing fundamental bounds on the achievable fidelity with which a desired operator can be synthesized in a given real-world reconfigurable wave system would allow for a number of definitive conclusions. For instance, if the upper bound on the achievable fidelity is below the minimum desired level of fidelity, then one can immediately conclude that it is impossible to tune the given system to the desired functionality, irrespective of the utilized optimization technique. In this way, one can screen candidate architectures for system embodiment to exclude those for which the bound certifies the infeasibility of reaching a desired fidelity. 
In addition, if a simple optimization technique almost reaches the upper bound, then one can immediately conclude that there cannot be other optimization techniques achieving substantially better outcomes. 
To date, it remains (to the best of our knowledge) unknown how to derive fundamental bounds on the achievable fidelity with which a desired operator can be synthesized in a real-world reconfigurable wave system. To be of practical value in the real-world, the derivation of such bounds should account for all relevant specifics of a given reconfigurable wave system, including electromagnetic interactions between tunable elements, as well as the limited programmability and attenuation of the tunable elements. 

In this paper, we consider the broad class of linear wave systems whose reconfigurability stems from tunable lumped elements. Tunable lumped elements such as PIN diodes are the predominant form of wave-domain programmability in contemporary wave systems involving programmable metasurfaces, such as reconfigurable intelligent surfaces (RISs) and dynamic metasurface antennas (DMAs). Our starting point is a physics-consistent system model based on multiport-network theory (MNT) whose parameters can be estimated experimentally to minimize model-reality mismatch. To bound the achievable operator-synthesis fidelity, we formulate the problem as a quadratically constrained quadratic program (QCQP), we relax the QCQP via semidefinite relaxation (SDR) to a convex semidefinite program (SDP), and we solve the SDP with a convex solver. Because all feasible points of the original QCQP are also feasible for the relaxed SDP, the SDP solution is an upper bound to the original QCQP. 

Our paper is organized as follows. In Sec.~\ref{sec_SystemModel}, we describe our system model. In Sec.~\ref{sec_FrobeniusBounds}, we consider the simpler problem of bounding the achievable Frobenius norm of the end-to-end MIMO transfer function, which is relevant to multiple recent works on RISs. Then, in Sec.~\ref{sec_FidelityBounds}, we bound the achievable fidelity; we reformulate the fractional-quadratic objective for fidelity maximization as a quadratic objective with an additional scaling variable, and we reuse the quadratic constraints formulated in Sec.~\ref{sec_FrobeniusBounds}. Next, we apply our bounds to three distinct experimental setups in Sec.~\ref{sec_Experimental}. We close with a brief discussion in Sec.~\ref{sec_discussion} and a conclusion in Sec.~\ref{sec_conclusion}.

Our main contributions are summarized as follows. \textit{First}, we derive fundamental physics-consistent bounds on Frobenius-norm maximization and fidelity maximization in real-world linear MIMO systems parametrized by 1-bit tunable lumped elements. \textit{Second}, we evaluate these bounds for three distinct experimental systems based on RISs and DMAs. This evaluation of our bounds for real-world experimental systems with unknown geometry and material composition is an important feature of our work. \textit{Third}, we systematically study how our bounds vary with the number of tunable elements, and we probe the bounds' tightness by trying to approach them with various discrete-optimization techniques.

Our work contributes to the development of an electromagnetic information theory (EIT) for \textit{programmable} wave systems~\cite{di2024electromagnetic}. The most closely related work is~\cite{salmi2026electromagnetically}. Therein, a similar SDR-based strategy was used to bound single-input-single-output (SISO) channel gain maximization, which is a special case of the Frobenius-norm maximization that we consider in Sec.~\ref{sec_FrobeniusBounds}. Most existing works on EIT focus on effective electromagnetic degrees of freedom (EEMDOFs) in \textit{static} MIMO systems~\cite{franceschetti2009capacity,miller2019waves,shiu2000fading,miller2000communicating,piestun2000electromagnetic,poon2005degrees,migliore2006role,muharemovic2008antenna,yuan2021electromagnetic,pizzo2022landau,yuan2023effects,ruiz2023degrees}. Some works in this space treat RISs as holographic transmitters or receivers capable of generating arbitrary aperture fields~\cite{dardari_JSAC,decarli2021communication}. Only a few works treat RISs as \textit{programmable} scatterers, e.g., to formulate EIT-based optimization objectives related to maximizing the number of EEMDOFs~\cite{del2019optimally,del2019optimized,do2022line,ruizWSA}. Moreover,~\cite{del2025effective} examined the EEMDOFs relevant to backscatter communications, where information is encoded into the RIS configuration rather than into the input wavefront. In parallel, the antenna community has explored QCQP formulations for pattern synthesis with reactively loaded antenna arrays in~\cite{corcoles2015reactively,salmi2025optimization}.
In addition, QCQP formulations combined with SDR also play an important role in recent theoretical works on bounding attainable performances of inverse-designed nanophotonic devices~\cite{molesky2020hierarchical,kuang2020computational,gustafsson2020upper,liska2021fundamental,chao2022physical,shim2024fundamental,amaolo2024maximum,gertler2025many,virally2025many}. 
Altogether, the combination of the objectives tackled in the present work with our real-life feasibility constraints have not been considered before. Importantly, we emphasize that (with the exception of~\cite{salmi2026electromagnetically}) none of these related works has applied the derived bounds to real-world experimental systems.

\textit{Notation:}
$\mathbb{R}$, $\mathbb{C}$, and $\mathbb{B}\triangleq\{0,1\}$ denote the sets of real, complex, and binary numbers, respectively.
$\jmath\triangleq\sqrt{-1}$ denotes the imaginary unit.
$(\cdot)^*$ denotes elementwise complex conjugation.
$(\cdot)^\top$ and $(\cdot)^\dagger$ denote transpose and conjugate transpose, respectively.
$\Re\{\cdot\}$ denotes real part.
$|\cdot|$ and $\|\cdot\|_F$ denote absolute value and Frobenius norm, respectively.
$\|\cdot\|_2$ denotes Euclidean norm for vectors and spectral norm for matrices.
$\mathbf{I}_a$ denotes the $a\times a$ identity matrix.
$\mathbf 0$ and $\mathbf 1$ denote the all-zeros and all-ones vectors/matrices of appropriate sizes.
$\mathrm{diag}(\mathbf{a})$ denotes the diagonal matrix whose diagonal entries are given by the vector $\mathbf{a}$.
$A_{i,j}$ denotes the $(i,j)$th entry of matrix $\mathbf{A}$.
$\mathrm{tr}(\cdot)$ and $\mathrm{vec}(\cdot)$ denote the trace and column-stacking vectorization operators, respectively. $\mathrm{unvec}(\cdot)$ denotes the inverse of $\mathrm{vec}(\cdot)$, reshaping a length-$mn$ vector into an $m\times n$ matrix (with dimensions clear from context).
$\otimes$ denotes the Kronecker product.
$\mathbf{A}\succeq\mathbf 0$ and $\mathbf{A}\succ\mathbf 0$ denote positive semidefinite and positive definite matrices, respectively.
$\mathbf{A}_{\mathcal{B}\mathcal{C}}$ denotes the submatrix (block) of $\mathbf{A}$ selected by row indices $\mathcal{B}$ and column indices $\mathcal{C}$.

\section{System Model}
\label{sec_SystemModel}

In this section, we describe how our system model maps the binary control vector determining the configuration of the $N_\mathrm{S}$ 1-bit-programmable elements to the system's corresponding $N_\mathrm{R}\times N_\mathrm{T}$ end-to-end multiple-input-multiple-output (MIMO) transfer function. We proceed in two steps. \textit{First}, we explain the encoding of the binary control vector in the tunable elements' load vector that describes their physical scattering properties. \textit{Second}, we describe using MNT the mapping from load vector to end-to-end MIMO transfer function. Finally, at the end of this section, we briefly discuss why the model parameters need to be estimated experimentally, why that leads to inevitable parameter ambiguities, and why these ambiguities are operationally irrelevant.

The overarching approach consists in partitioning the reconfigurable system into three entities: (i) ports via which waves enter and/or exit; (ii) tunable lumped elements; (iii) static scattering objects. The tunable lumped elements are then described as ``virtual'' additional ports that are terminated by tunable loads. This system model applies to any linear wave system whose reconfigurability is based on tunable lumped elements. In particular, it applies to the experiments considered in this work: two distinct RIS-parametrized MIMO channels and a multi-feed DMA. The correspondence is as follows: the ports of the $N_\mathrm{T}$ transmitting and $N_\mathrm{R}$ receiving antennas in the RIS scenario play the same role as the $N_\mathrm{T}$ feeds and $N_\mathrm{R}$ user antenna ports in the DMA scenario. In both cases, the $N_\mathrm{R}\times N_\mathrm{T}$ end-to-end MIMO transfer function is parametrized by $N_\mathrm{S}$ tunable lumped elements.

\textit{Encoding Function:} The configuration of the $N_\mathrm{S}$ 1-bit-programmable lumped elements is described by a control vector $\mathbf{v}\in\mathbb{B}^{N_\mathrm{S}}$. Because within MNT we model each tunable lumped element as a ``virtual'' port terminated by a tunable load, the physical scattering characteristics of the $i$th tunable element are captured by the reflection coefficient $r_i\in\mathbb{C}$ of the associated $i$th load. Assuming that all $N_\mathrm{S}$ elements are identical and independently configurable, the mapping from the control vector $\mathbf{v}$ to the load vector $\mathbf{r}\in\mathbb{C}^{N_\mathrm{S}}$ is affine\footnote{For multi-bit-programmable elements, the encoding function is generally not affine; if the tunable elements cannot be configured independently (e.g., due to limited power supply for the control circuit), the encoding function can be very complex.~\cite{ContRIS_LWC,largeRIS_TCOM,del2025experimentalreducedrank,pWDCperspective}}~\cite{ContRIS_LWC,largeRIS_TCOM,del2025experimentalreducedrank,pWDCperspective}:
\begin{equation}
    \mathbf{r}(\mathbf{v}) = \alpha \,\mathbf{1} + (\beta-\alpha)\, \mathbf{v},
    \label{eq_encod}
\end{equation}
where $\alpha,\beta\in\mathbb{C}$ are the two possible reflection coefficients.

\textit{MNT Function:}
Given the aforementioned partition of the reconfigurable system into three entities, the third entity comprising all static scattering objects can be described as a static system with $N=N_\mathrm{T}+N_\mathrm{R}+N_\mathrm{S}$ ports that is fully characterized by its scattering matrix $\mathbf{S}\in\mathbb{C}^{N\times N}$. In addition, the ensemble of $N_\mathrm{S}$ tunable loads can be interpreted as an $N_\mathrm{S}$-port system whose scattering matrix is $\mathbf{\Phi}(\mathbf{r})=\mathrm{diag}(\mathbf{r})$. We define $\mathbf{S}$ and $\mathbf{\Phi}$ using the same reference impedance $Z_0=50\ \Omega$ at all ports.
The ``virtual'' ports of the static $N$-port system are connected to the $N_\mathrm{S}$-port ensemble of tunable loads. Standard MNT yields the mapping from $\mathbf{r}$ to the resulting end-to-end MIMO transfer matrix $\mathbf{H}\in\mathbb{C}^{N_\mathrm{R}\times N_\mathrm{T}}$~\cite{matteo_universal,ContRIS_LWC,largeRIS_TCOM,del2025experimentalreducedrank,pWDCperspective}: 
\begin{equation}
\mathbf{H}(\mathbf{r}) = \mathbf{H}_0 + \mathbf{A}\,\bigl(\mathbf{I}_{N_\mathrm{S}}\,-\,\mathbf{\Phi}(\mathbf{r})\,\mathbf{\Gamma}\bigr)^{-1}\,\mathbf{\Phi}(\mathbf{r})\,\mathbf{B},
\label{eq_MNT}
\end{equation}
where, for notational ease, $\mathbf{H}_0 \triangleq \mathbf{S}_\mathcal{RT}\in\mathbb{C}^{N_\mathrm{R}\times N_\mathrm{T}}$, $\mathbf{A} \triangleq \mathbf{S}_\mathcal{RS}\in\mathbb{C}^{N_\mathrm{R}\times N_\mathrm{S}}$, $\mathbf{\Gamma} \triangleq \mathbf{S}_\mathcal{SS}\in\mathbb{C}^{N_\mathrm{S}\times N_\mathrm{S}}$, and $\mathbf{B} \triangleq \mathbf{S}_\mathcal{ST}\in\mathbb{C}^{N_\mathrm{S}\times N_\mathrm{T}}$, and  $\mathcal{T}$, $\mathcal{R}$, and $\mathcal{S}$ denote the sets containing the port indices associated with the transmitting ports, the receiving ports, and ``virtual'' ports, respectively.

\textit{Parameter Ambiguities:}
The parameters required by our system model to map $\mathbf{v}\rightarrow \mathbf{r} \rightarrow \mathbf{H}$ are $\alpha$ and $\beta$ for the encoding function and $\mathbf{H}_0$, $\mathbf{A}$, $\mathbf{\Gamma}$, and $\mathbf{B}$ for the MNT function. If all system details (geometry and material composition) are perfectly known, and if computational complexity is not a limiting factor, then all MNT parameters can be obtained with a single full-wave simulation~\cite{tapie2023systematic,zheng2024mutual,almunif2025network}. However, for real-world experiments, these conditions are rarely met. Thus, it is necessary to estimate the parameters experimentally. Unfortunately, the MNT parameters cannot be measured directly in most cases because the ``virtual'' ports are usually not connectorized and too numerous\footnote{A typical vector network analyzer has less than 10 ports, while $N_\mathrm{S}$ is typically at least on the order of 100.}. Thus, the model parameters must be estimated indirectly based on experimental measurements.
If the encoding function is perfectly known \textit{and} satisfies certain criteria\footnote{At least three distinct terminations are required for each ``virtual'' port, and it is necessary to connect pairs of neighboring ``virtual'' ports (as well as at least one pair comprising an actual and a ``virtual'' port) via coupled loads.~\cite{del2024virtual2p0,del2025virtual3p0,del2025wireless}}, then the MNT parameters can be experimentally estimated without any ambiguity~\cite{del2024virtual2p0,del2025virtual3p0}. However, in typical real-world reconfigurable systems such as those considered in this work, the encoding function does not satisfy these criteria (only two individual loads are available) and its parameters ($\alpha$ and $\beta$) are not perfectly known. Nonetheless, various techniques for experimentally estimating \textit{proxy} parameter sets exist~\cite{sol2024experimentally,ContRIS_LWC,largeRIS_TCOM,del2025experimentalreducedrank,tapie2025experimental}. These proxy parameters accurately map any control vector to the corresponding end-to-end MIMO transfer function. Thus even though the parameter values are ambiguous, using the system model with proxy parameters is operationally equivalent (in terms of physically observable quantities) to using the system model with the true parameters. In addition, the parameter ambiguities do not affect SDR-based bounds; the argument to this effect for SISO systems in the Appendix of~\cite{salmi2026electromagnetically} carries directly over to MIMO systems. Because these ambiguities are operationally irrelevant within the scope of the present paper, we do not detail them further here; interested readers are referred to Sec.~III in~\cite{salmi2026electromagnetically} for an overview.

\section{SDR-Based Bound on the Achievable Frobenius Norm of the MIMO Transfer Function}
\label{sec_FrobeniusBounds}

As a preliminary step, we derive in this section an upper bound on the Frobenius norm of the end-to-end MIMO transfer function. This constitutes a generalization of the bound on the SISO channel gain in~\cite{salmi2026electromagnetically} to multiport excitation. The next section will build on the quadratic formulations of objective and constraints derived in this section. 

Physically, the squared Frobenius norm represents the aggregate power gain across all spatial modes of the MIMO system (assuming identity covariance of excitation signals). Optimizing the RIS configuration to maximize the Frobenius norm of a MIMO transfer function can thus be interpreted as shaping the MIMO channel for maximum power gain. This optimization goal was pursued in~[(39),~\cite{TSP_MIMOChannelShaping}] as part of a low-complexity solution for a MIMO rate-maximization problem. Optimization goals related to maximizing the Frobenius norm were also previously studied in~\cite{ICC_TraceMax,SPL_SNR}. While~\cite{ICC_TraceMax,SPL_SNR,TSP_MIMOChannelShaping} state that they identify globally optimal configurations, they neglect electromagnetic effects such as mutual coupling and practical hardware constraints such as few-bit-programmability and attenuation of the tunable loads. Moreover,~\cite{ICC_TraceMax,SPL_SNR,TSP_MIMOChannelShaping} assume that the load network connected to the ``virtual'' RIS ports is ``beyond-diagonal'' -- see our related discussion in Sec.~\ref{sec_discussion}. Thus, although~\cite{ICC_TraceMax,SPL_SNR,TSP_MIMOChannelShaping} derive optimal solutions under simplifying and idealized assumptions, they do not yield an electromagnetically consistent upper bound on the Frobenius norm of a real-world RIS-parametrized MIMO channel. We derive such a bound in this section. Beyond constituting a preliminary step for the next section, the results of this section are thus themselves of practical relevance in wireless communications.

Formally, the Frobenius norm maximization problem considered in this section is summarized as follows:
\begin{equation}
\begin{aligned}
\max_{\mathbf r\in\mathbb C^{N_\mathrm S}}\quad 
& \|\mathbf H(\mathbf r)\|_F^2\\
\text{s.t.}\quad 
& r_i\in\{\alpha,\beta\},\quad i=1,\dots,N_\mathrm S.
\end{aligned}
\label{eq:frob_opt}
\end{equation}
The functional dependence of $\mathbf{H}$ on $\mathbf{r}$ is given in (\ref{eq_MNT}).

We begin by defining the auxiliary matrix variable
\begin{equation}
\mathbf X \triangleq \bigl(\mathbf{I}_{N_\mathrm{S}} -\mathbf\Phi(\mathbf{r})\,\mathbf\Gamma\bigr)^{-1}\mathbf\Phi(\mathbf{r})\,\mathbf B \in \mathbb{C}^{N_\mathrm{S} \times N_\mathrm{T}}
\label{eq_def_X}
\end{equation}
so that we can rewrite (\ref{eq_MNT}) as
\begin{equation}
\mathbf{H}(\mathbf{r})=\mathbf{H}_0+\mathbf A\,\mathbf X.
\label{eq:eqn_w_x}
\end{equation}

\textit{Quadratic Objective:}
Now, we can express our objective from (\ref{eq:frob_opt}) in a form that is quadratic in $\mathbf{X}$:
\begin{align}
\|\mathbf H(\mathbf r)\|_F^2
&= \|\mathbf H_0+\mathbf A\,\mathbf X\|_F^2 \nonumber\\
&= \mathrm{tr}\Big((\mathbf H_0+\mathbf A\,\mathbf X)^{\dagger}(\mathbf H_0+\mathbf A\,\mathbf X)\Big) \nonumber\\
&= \mathrm{tr}(\mathbf X^{\dagger}\,\mathbf A^{\dagger}\,\mathbf A\,\mathbf X)
+2\,\Re\!\left\{\mathrm{tr}(\mathbf X^{\dagger}\,\mathbf A^{\dagger}\,\mathbf H_0)\right\}
+\|\mathbf H_0\|_F^2 .
\label{eq:frob_expand}
\end{align}
To work with an optimization variable in vector form rather than in matrix form, we define
\begin{equation}
   \mathbf y\triangleq \mathrm{vec}(\mathbf X)\in\mathbb C^{N_\mathrm S N_\mathrm T}. 
\label{eq_def_y}
\end{equation}
Using various identities,\footnote{We use the following three matrix identities~\cite{PetersenPedersenCookbook}:\begin{enumerate}[label=(\arabic*), leftmargin=*, itemsep=0pt, topsep=2pt]
\item $\mathrm{vec}(\mathbf{\Theta}\,\mathbf{\Lambda}\,\mathbf{\Xi}) = (\mathbf{\Xi}^{\top}\!\otimes\!\mathbf{\Theta})\,\mathrm{vec}(\mathbf{\Lambda})$.
\item $\mathrm{tr}(\mathbf{\Lambda}^{\dagger}\,\mathbf{\Psi}\,\mathbf{\Lambda})=\mathrm{vec}(\mathbf{\Lambda})^{\dagger}\,(\mathbf{I}\otimes \mathbf{\Psi})\,\mathrm{vec}(\mathbf{\Lambda})$.
\item $\mathrm{tr}(\mathbf{\Lambda}^{\dagger}\,\mathbf{\Omega})=\mathrm{vec}(\mathbf{\Omega})^{\dagger}\,\mathrm{vec}(\mathbf{\Lambda})$.
\end{enumerate}}
we can rewrite (\ref{eq:frob_expand}) in terms of $\mathbf{y}$:
\begin{equation}
\|\mathbf H(\mathbf r)\|_F^2
= \mathbf y^{\dagger}\,\mathbf R_0\,\mathbf y
+2\,\Re\!\left\{\mathbf q_0^{\dagger}\,\mathbf y\right\}
+\tau_0,
\label{eq:frob_quad_obj}
\end{equation}
where $\mathbf R_0 \triangleq \mathbf I_{N_\mathrm T}\otimes (\mathbf A^{\dagger}\,\mathbf A)$, $\mathbf q_0 \triangleq \mathrm{vec}(\mathbf A^{\dagger}\,\mathbf H_0)$, and $\tau_0 \triangleq \|\mathbf H_0\|_F^2$.

\textit{Quadratic Binary Constraints:}
Next, we formulate our constraints in a form that is quadratic in $\mathbf y$. In the original formulation of our optimization problem in (\ref{eq:frob_opt}), we had $N_\mathrm{S}$ 1-bit programmability constraints (one for each RIS element). Now, however, we are working with the auxiliary optimization variable $\mathbf{y}$ which has $N_\mathrm{S}N_\mathrm{T}$ entries that are related to $\mathbf{r}$ in a complicated manner via (\ref{eq_def_y}) and (\ref{eq_def_X}). 

When $\mathbf{r}$ is the optimization variable, the logical OR condition associated with the 1-bit-programmability constraint of the $i$th tunable element can be expressed as $(r_i-\alpha)^*(r_i-\beta)=0 \ \forall \ i$~\cite{salmi2026electromagnetically,shim2024fundamental,gertler2025many}. To express the constraint in terms of $\mathbf{X}$, we first note that (\ref{eq_def_X}) implies  $\bigl(\mathbf{I}_{N_\mathrm{S}} -\mathbf\Phi(\mathbf{r})\,\mathbf\Gamma\bigr) \,\mathbf X = \mathbf\Phi(\mathbf{r})\,\mathbf B$. Rearranging yields
\begin{equation}
    \mathbf{X} = \mathbf{\Phi}(\mathbf{r})\, \mathbf{Z},
    \label{eqXZ}
\end{equation}
where $\mathbf{Z}\triangleq \mathbf{B}+\mathbf{\Gamma}\mathbf{X}$. This matrix equality is equivalent to $N_\mathrm{S}N_\mathrm{T}$ scalar equalities of the form $X_{s,t}=r_s Z_{s,t}$ for the $(s,t)$th entry. Because $r_s$ has only two possible values ($\alpha$ or $\beta$), it follows that $X_{s,t} = \alpha Z_{s,t}$ or $X_{s,t} = \beta Z_{s,t}$. Following~\cite{salmi2026electromagnetically,shim2024fundamental,gertler2025many}, this logical OR condition can be expressed as
\begin{equation}
(X_{s,t} - \alpha Z_{s,t})^*(X_{s,t} - \beta Z_{s,t})=0.
\label{eq32}
\end{equation}

To recast (\ref{eq32}) into a quadratic form in our auxiliary vector variable $\mathbf y$ defined in (\ref{eq_def_y}), we now express both $X_{s,t}$ and $Z_{s,t}$ as affine functions of $\mathbf y$. Let $\mathbf{u}_s\in\mathbb{C}^{N_\mathrm{S}}$ be the $s$th canonical basis vector in $\mathbb{C}^{N_\mathrm{S}}$, and let  $\mathbf{w}_t\in\mathbb{C}^{N_\mathrm{T}}$ be the $t$th canonical basis vector in $\mathbb{C}^{N_\mathrm{T}}$. Then,
\begin{equation}
X_{s,t}
= \mathbf u_s^{\top}\mathbf X\,\mathbf w_t
= \bigl(\mathbf w_t^{\top}\!\otimes \mathbf u_s^{\top}\bigr)\,\mathrm{vec}(\mathbf X)
= \mathbf{e}_{s,t}^{\top}\,\mathbf y,
\label{eq_Xst}
\end{equation}
where $\mathbf{e}_{s,t} \triangleq \mathbf w_t \otimes \mathbf u_s\in\mathbb C^{N_\mathrm S N_\mathrm T}$. Similarly, we have
\begin{equation}
\begin{aligned}
Z_{s,t}
&=  \mathbf u_s^{\top}\left(\mathbf{B}+\mathbf{\Gamma}\mathbf{X} \right)\mathbf w_t \\
&= B_{s,t} + \bigl(\mathbf w_t^{\top}\!\otimes (\mathbf u_s^{\top}\mathbf{\Gamma})\bigr)\,\mathrm{vec}(\mathbf X) \\
&= B_{s,t} + \mathbf f_{s,t}^{\top}\mathbf y .
\end{aligned}
\label{eq_Zst}
\end{equation}
where $\mathbf{f}_{s,t}\triangleq \mathbf w_t \otimes (\mathbf{\Gamma}^\top \mathbf u_s)\in\mathbb C^{N_\mathrm S N_\mathrm T}$.
Substituting (\ref{eq_Xst}) and (\ref{eq_Zst}) into (\ref{eq32}) yields
\begin{equation}
\bigl(\left( \mathbf e_{s,t}-\alpha\,\mathbf f_{s,t} \right)^{\top}\mathbf y -\alpha B_{s,t}\bigr)^*
\bigl(\left(\mathbf e_{s,t}-\beta\,\mathbf f_{s,t} \right)^{\top}\mathbf y  -\beta B_{s,t}\bigr)=0.
\label{eq:mimo_affprod}
\end{equation}
Expanding (\ref{eq:mimo_affprod}) yields the sought-after quadratic equality for $\mathbf{y}$:
\begin{equation}
\mathbf y^\dagger \mathbf R_{s,t}\mathbf y
+\mathbf y^\dagger \mathbf p_{s,t}
+\mathbf q_{s,t}^\dagger \mathbf y
+\tau_{s,t}=0,
\label{eq_frob_binary_constraint}
\end{equation}
where 
\begin{equation}
\begin{aligned}
\mathbf R_{s,t}
&\triangleq \bigl(\mathbf e_{s,t}-\alpha\,\mathbf f_{s,t}\bigr)^*
            \bigl(\mathbf e_{s,t}-\beta\,\mathbf f_{s,t}\bigr)^{\top}, \nonumber\\
\mathbf p_{s,t}
&\triangleq -\beta B_{s,t}\,\bigl(\mathbf e_{s,t}-\alpha\,\mathbf f_{s,t}\bigr)^*, \nonumber\\
\mathbf q_{s,t}
&\triangleq -\alpha B_{s,t}\,\bigl(\mathbf e_{s,t}-\beta\,\mathbf f_{s,t}\bigr)^*, \nonumber\\
\tau_{s,t}
&\triangleq \alpha^*\beta\,|B_{s,t}|^2.
\end{aligned}
\end{equation}

\textit{Quadratic Repetition Constraints:}
The quadratic equality constraint in (\ref{eq_frob_binary_constraint}) specializes for $N_\mathrm{T}=1$ to the 1-bit-programmability constraints used in the SISO case in~\cite{salmi2026electromagnetically}. However, for $N_\mathrm{T}>1$, (\ref{eq_frob_binary_constraint}) enforces that, for each pair $(s,t)$, the relation
$X_{s,t}=\alpha Z_{s,t}$ OR $X_{s,t}=\beta Z_{s,t}$ holds, but it does \emph{not}
enforce that the \emph{same} choice ($\alpha$ versus $\beta$) is used for a given
element $s$ across all $t=1,\dots,N_\mathrm{T}$. Therefore, in the case of $N_\mathrm{T}>1$ that is considered in the present paper, we need to formulate additional ``repetition constraints''; related repetition constraints are discussed in~\cite{shim2024fundamental,salmi2025optimization}.

We can formulate our repetition constraints based on (\ref{eq32}) by enforcing that, for each index $s$, the same binary choice ($\alpha$ or $\beta$) is taken for all indices $t=1,\dots,N_\mathrm{T}$. To this end, we select an arbitrary reference index $t_0$ (without loss of generality, $t_0\triangleq 1$). Then, for each $t\neq t_0$, we must impose additional constraints such that $X_{s,t}=\alpha\,Z_{s,t}$ whenever $X_{s,t_0}=\alpha\,Z_{s,t_0}$ and $X_{s,t}=\beta\,Z_{s,t}$ whenever $X_{s,t_0}=\beta\,Z_{s,t_0}$.
Following the logical-OR structure underlying (\ref{eq32}), we can rule out that $(s,t)$ selects $\alpha$ while $(s,t_0)$ selects $\beta$, or vice versa, with the following pair of constraints for all $s\in\{1,\dots,N_\mathrm{S}\}$ and all $t\in\{1,\dots,N_\mathrm{T}\}\setminus\{t_0\}$:
\begin{subequations}
\begin{equation}
\bigl(X_{s,t}-\alpha Z_{s,t}\bigr)^*\bigl(X_{s,t_0}-\beta Z_{s,t_0}\bigr) = 0,
\label{eq_rep1}
\end{equation}
\begin{equation}
\bigl(X_{s,t}-\beta Z_{s,t}\bigr)^*\bigl(X_{s,t_0}-\alpha Z_{s,t_0}\bigr) = 0.
\label{eq_rep2}
\end{equation}
\label{eq_repetitionconstraint}
\end{subequations}
To express (\ref{eq_repetitionconstraint}) in terms of $\mathbf y$, we substitute (\ref{eq_Xst}) and (\ref{eq_Zst}) into (\ref{eq_repetitionconstraint}) and obtain
\begin{subequations}
\begin{equation}
\bigl(\left(\mathbf e_{s,t}-\alpha \mathbf f_{s,t}\right)^{\top}\mathbf y-\alpha B_{s,t}\bigr)^*
\bigl(\left(\mathbf e_{s,t_0}-\beta \mathbf f_{s,t_0}\right)^{\top}\mathbf y-\beta B_{s,t_0}\bigr)=0,
\label{eq_rep1_affprod}
\end{equation}
\begin{equation}
\bigl(\left(\mathbf e_{s,t}-\beta \mathbf f_{s,t}\right)^{\top}\mathbf y-\beta B_{s,t}\bigr)^*
\bigl(\left(\mathbf e_{s,t_0}-\alpha \mathbf f_{s,t_0}\right)^{\top}\mathbf y-\alpha B_{s,t_0}\bigr)=0.
\label{eq_rep2_affprod}
\end{equation}
\end{subequations}
Expanding (\ref{eq_rep1_affprod}) and (\ref{eq_rep2_affprod}) yields two quadratic equalities in $\mathbf y$:
\begin{subequations}
\begin{equation}
\mathbf y^\dagger \mathbf R^{(1)}_{s,t}\mathbf y
+\mathbf y^\dagger \mathbf p^{(1)}_{s,t}
+\bigl(\mathbf q^{(1)}_{s,t}\bigr)^\dagger \mathbf y
+\tau^{(1)}_{s,t}=0,
\label{eq_rep1_qcqp}
\end{equation}
\begin{equation}
\mathbf y^\dagger \mathbf R^{(2)}_{s,t}\mathbf y
+\mathbf y^\dagger \mathbf p^{(2)}_{s,t}
+\bigl(\mathbf q^{(2)}_{s,t}\bigr)^\dagger \mathbf y
+\tau^{(2)}_{s,t}=0,
\label{eq_rep2_qcqp}
\end{equation}
\label{eq_rep_qcqp}
\end{subequations}
where
\begin{subequations}
\begin{align}
\mathbf R^{(i)}_{s,t}
&\triangleq \bigl(\mathbf e_{s,t}-\mu\,\mathbf f_{s,t}\bigr)^*\,
            \bigl(\mathbf e_{s,t_0}-\nu\,\mathbf f_{s,t_0}\bigr)^{\top},
\label{eq_Rrep}\\
\mathbf p^{(i)}_{s,t}
&\triangleq -\nu B_{s,t_0}\,\bigl(\mathbf e_{s,t}-\mu\,\mathbf f_{s,t}\bigr)^*,
\label{eq_prep}\\
\mathbf q^{(i)}_{s,t}
&\triangleq -\mu\, B_{s,t}\,\bigl(\mathbf e_{s,t_0}-\nu\,\mathbf f_{s,t_0}\bigr)^*,
\label{eq_qrep}\\
\tau^{(i)}_{s,t}
&\triangleq \mu^*\nu\, B_{s,t}^* B_{s,t_0},
\label{eq_taurep}
\end{align}
\end{subequations}
with $\mu=\alpha$ and $\nu=\beta$ for $i=1$ while $\mu=\beta$ and $\nu=\alpha$ for $i=2$.

\textit{QCQP Formulation:}
Combining the quadratic objective in (\ref{eq:frob_quad_obj}) with the $N_\mathrm{S}N_\mathrm{T}$ quadratic 1-bit-programmability constraints in (\ref{eq_frob_binary_constraint}) and the $2N_\mathrm{S}(N_\mathrm{T}-1)$ repetition constraints in (\ref{eq_rep_qcqp}) yields the following quadratically constrained quadratic program (QCQP):
\begin{equation}
\small
\begin{aligned}
\max_{\mathbf y\in\mathbb C^{N_\mathrm{S}N_\mathrm{T}}}\quad
& \mathbf y^{\dagger}\mathbf R_0 \mathbf y
+2\,\Re\!\left\{\mathbf q_0^{\dagger}\mathbf y\right\}
+\tau_0\\
\text{s.t.}\quad
& \mathbf y^\dagger \mathbf R_{s,t}\mathbf y
+\mathbf y^\dagger \mathbf p_{s,t}
+\mathbf q_{s,t}^\dagger \mathbf y
+\tau_{s,t}=0,\\
& \hspace{1.7em}\substack{
\forall\, s\in\{1,\dots,N_\mathrm S\},\\
\forall\, t\in\{1,\dots,N_\mathrm T\},
}\\[0.2em]
& \mathbf y^\dagger \mathbf R^{(i)}_{s,t}\mathbf y
+\mathbf y^\dagger \mathbf p^{(i)}_{s,t}
+\bigl(\mathbf q^{(i)}_{s,t}\bigr)^\dagger \mathbf y
+\tau^{(i)}_{s,t}=0,\\
& \hspace{1.7em}\substack{
\forall\, s\in\{1,\dots,N_\mathrm S\},\\
\forall\, t\in\{2,\dots,N_\mathrm T\},\\
\forall\, i\in\{1,2\}.
}
\end{aligned}
\label{eq:frob_qcqp}
\end{equation}
\normalsize

Now, we rewrite the quadratic terms so that they become linear in a higher-dimensional matrix variable via the standard lifting technique. We introduce
\begin{equation}
\mathbf Y \triangleq \mathbf y\mathbf y^\dagger \in \mathbb C^{N_\mathrm S N_\mathrm T \times N_\mathrm S N_\mathrm T},
\label{eq:Y_def}
\end{equation}
which is Hermitian positive semidefinite and, by construction, satisfies $\mathrm{rank}(\mathbf Y)=1$.
Using the property $\mathbf y^\dagger \mathbf R\,\mathbf y=\mathrm{tr}(\mathbf R\,\mathbf y\mathbf y^\dagger)=\mathrm{tr}(\mathbf R\,\mathbf Y)$, we can rewrite the QCQP in \eqref{eq:frob_qcqp} equivalently as
\begin{equation}
\small
\begin{aligned}
\max_{\mathbf y,\mathbf Y}\quad
& \mathrm{tr}(\mathbf R_0\mathbf Y)
+2\,\Re\!\left\{\mathbf q_0^{\dagger}\mathbf y\right\}
+\tau_0\\
\text{s.t.}\quad
& \mathrm{tr}(\mathbf R_{s,t}\mathbf Y)
+\mathbf y^\dagger \mathbf p_{s,t}
+\mathbf q_{s,t}^\dagger \mathbf y
+\tau_{s,t}=0,\\
& \hspace{1.7em}\substack{
\forall\, s\in\{1,\dots,N_\mathrm S\},\\
\forall\, t\in\{1,\dots,N_\mathrm T\},
}\\[0.2em]
& \mathrm{tr}(\mathbf R^{(i)}_{s,t}\mathbf Y)
+\mathbf y^\dagger \mathbf p^{(i)}_{s,t}
+\bigl(\mathbf q^{(i)}_{s,t}\bigr)^\dagger \mathbf y
+\tau^{(i)}_{s,t}=0,\\
& \hspace{1.7em}\substack{
\forall\, s\in\{1,\dots,N_\mathrm S\},\\
\forall\, t\in\{2,\dots,N_\mathrm T\},\\
\forall\, i\in\{1,2\},
}\\[0.2em]
& \mathbf Y=\mathbf y\mathbf y^\dagger.
\end{aligned}
\label{eq:qcqp_lifted}
\end{equation}
\normalsize

\textit{Semidefinite Relaxation:}
Since the rank-one constraint is the only non-convex part of the lifted QCQP formulation in \eqref{eq:qcqp_lifted}, we obtain a convex semidefinite program (SDP) by dropping it.
This constitutes the semidefinite relaxation (SDR) of our QCQP.
Specifically, we relax $\mathbf Y=\mathbf y\mathbf y^\dagger$ to $\mathbf Y\succeq \mathbf y\mathbf y^\dagger$.
To express this condition as a linear matrix inequality, we consider the block matrix $\mathbf M \triangleq \begin{bmatrix}\mathbf Y & \mathbf y\\ \mathbf y^\dagger & 1\end{bmatrix}$.
Since the bottom-right block is positive, the Schur complement identity implies
$\mathbf M \succeq \mathbf 0  \Longleftrightarrow  \mathbf Y-\mathbf y\mathbf y^\dagger \succeq \mathbf 0$, i.e., $\mathbf M\succeq \mathbf 0$ is a linear matrix inequality representing $\mathbf Y \succeq \mathbf y\mathbf y^\dagger$.
The SDR of the QCQP in \eqref{eq:qcqp_lifted} thus becomes the following SDP:
\begin{equation}
\small
\begin{aligned}
\max_{\mathbf y,\mathbf Y}\quad
& \mathrm{tr}(\mathbf R_0\mathbf Y)
+2\,\Re\!\left\{\mathbf q_0^{\dagger}\mathbf y\right\}
+\tau_0\\
\text{s.t.}\quad
& \mathrm{tr}(\mathbf R_{s,t}\mathbf Y)
+\mathbf y^\dagger \mathbf p_{s,t}
+\mathbf q_{s,t}^\dagger \mathbf y
+\tau_{s,t}=0,\\
& \hspace{1.7em}\substack{
\forall\, s\in\{1,\dots,N_\mathrm S\},\\
\forall\, t\in\{1,\dots,N_\mathrm T\},
}\\[0.2em]
& \mathrm{tr}(\mathbf R^{(i)}_{s,t}\mathbf Y)
+\mathbf y^\dagger \mathbf p^{(i)}_{s,t}
+\bigl(\mathbf q^{(i)}_{s,t}\bigr)^\dagger \mathbf y
+\tau^{(i)}_{s,t}=0,\\
& \hspace{1.7em}\substack{
\forall\, s\in\{1,\dots,N_\mathrm S\},\\
\forall\, t\in\{2,\dots,N_\mathrm T\},\\
\forall\, i\in\{1,2\},
}\\[0.2em]
&
\begin{bmatrix}
\mathbf Y & \mathbf y\\
\mathbf y^\dagger & 1
\end{bmatrix}\succeq \mathbf 0.
\end{aligned}
\label{eq:sdr_sdp}
\end{equation}
\normalsize
Because any feasible point of our original QCQP in \eqref{eq:qcqp_lifted}
is also feasible for \eqref{eq:sdr_sdp}, the optimal value of \eqref{eq:sdr_sdp} upper-bounds the optimal value of \eqref{eq:frob_qcqp}.
The SDP can be solved with standard convex-optimization frameworks; in this work,
we use the CVX framework~\cite{cvx} with the SeDuMi solver.

Our SDR-based bound for the Frobenius norm of the end-to-end MIMO transfer function is defined as the optimal objective value of the relaxed SDP. Given a solver output $(\check{\mathbf Y},\check{\mathbf y})$ for \eqref{eq:sdr_sdp}, we define
\begin{equation}
B^\mathrm{Fro}_{\mathrm{SDR}}
\triangleq
\mathrm{tr}(\mathbf R_0\,\check{\mathbf Y})
+2\,\Re\!\left\{\mathbf q_0^{\dagger}\,\check{\mathbf y}\right\}
+\tau_0.
\label{eq:Bsdr_def}
\end{equation}

The derivation in the Appendix of~\cite{salmi2026electromagnetically} for the invariance of the SDR-based SISO channel gain bound with respect to parameter ambiguities extends directly to the present MIMO formulation and is therefore not repeated here. Indeed, each ambiguity induces an invertible affine change of variables on $\mathbf{y}$, which bijectively maps feasible sets of the relaxed SDP while preserving the objective value.

\textit{Norm-Inequality Benchmark Bound:}
To compute an electromagnetically consistent benchmark for our SDR-based bound in (\ref{eq:Bsdr_def}), we apply a norm-inequality argument to (\ref{eq_MNT}). 
Using various matrix identities,\footnote{We use the following four matrix identities~\cite{PetersenPedersenCookbook}:\begin{enumerate}[label=(\arabic*), leftmargin=*, itemsep=0pt, topsep=2pt]
\item $\|\mathbf{\Theta}+\mathbf{\Lambda}\|_{F}\le \|\mathbf{\Theta}\|_{F}+\|\mathbf{\Lambda}\|_{F}$.
\item $\|\mathbf{\Theta}\mathbf{\Lambda}\|_{F}\le \|\mathbf{\Theta}\|_{2}\,\|\mathbf{\Lambda}\|_{F}$.
\item $\|\mathbf{\Theta}\|_{2}\le \|\mathbf{\Theta}\|_{F}$.
\item $\|(\mathbf{I}-\mathbf{\Omega})^{-1}\|_{2}\le (1-\|\mathbf{\Omega}\|_{2})^{-1}\quad \text{if }\|\mathbf{\Omega}\|_{2}<1$.
\end{enumerate}}
we obtain the norm-inequality (NI) bound $\|\mathbf{H}(\mathbf{r})\|_F^2 \le  B^\mathrm{Fro}_\mathrm{NI} $ with
\begin{equation}
    B^\mathrm{Fro}_\mathrm{NI} = \big(\|\mathbf{H}_0\|_F + \|\mathbf{A}\|_2\,\frac{\gamma}{1-\gamma\|\mathbf{\Gamma}\|_2}\,\|\mathbf{B}\|_F\big)^2
\end{equation}
where $\gamma\triangleq\max\{|\alpha|,|\beta|\}$ and for which $1-\gamma\|\mathbf{\Gamma}\|_2>0$ is a sufficient validity condition. 
Unlike our SDR-based bound, this NI-based bound does depend on the gauge freedoms of the proxy parameters. Consequently, similarly to [(21),~\cite{salmi2026electromagnetically}], we can tighten the NI bound by optimizing the admissible gauge parameters, yielding an optimized norm-inequality bound (NIO). We implement this gauge optimization for the NIO bound via a gradient-based routine in TensorFlow that minimizes $B^\mathrm{Fro}_\mathrm{NI}$. We define $B^\mathrm{Fro}_\mathrm{NIO}$ as the minimized value of $B^\mathrm{Fro}_\mathrm{NI}$ obtained with the optimized gauge parameters.

\section{SDR-Bounds on the Realizable \\MIMO Transfer Function Fidelity}
\label{sec_FidelityBounds}

In this section, we derive an upper bound on the fidelity with which $\mathbf{H}(\mathbf{r})$ can approximate a desired linear operator $\mathbf{H}_\mathrm{des}$. We define the scale-invariant fidelity as 
\begin{equation}
\label{eq:fidelity_def}
F\!\left(\mathbf{H}(\mathbf{r}),\mathbf{H}_{\mathrm{des}}\right)
\triangleq
\frac{\left|\mathrm{tr}\!\left(\mathbf{H}_{\mathrm{des}}^{\dagger}\mathbf{H}(\mathbf{r})\right)\right|^{2}}
{\left\|\mathbf{H}_{\mathrm{des}}\right\|_{F}^{2}\,\left\|\mathbf{H}(\mathbf{r})\right\|_{F}^{2}}
\in[0,1].
\end{equation}
The numerator in \eqref{eq:fidelity_def} quantifies the coherent correlation between $\mathbf{H}(\mathbf{r})$ and $\mathbf{H}_{\mathrm{des}}$, while the denominator in \eqref{eq:fidelity_def} normalizes by their total strengths, making \eqref{eq:fidelity_def} independent of the overall strengths. Thereby, our fidelity metric is independent of the overall scaling and exclusively focuses on how well the shapes of $\mathbf{H}(\mathbf{r})$ and $\mathbf{H}_\mathrm{des}$ match. A unit fidelity hence implies that $\mathbf{H}(\mathbf{r})$ is proportional to $\mathbf{H}_\mathrm{des}$ up to a complex-valued global scaling factor: $\mathbf{H}(\mathbf{r})=a\mathbf{H}_\mathrm{des}$, where $a$ is a non-zero complex-valued scalar. 

The scale-invariance of our fidelity metric is deliberate for two reasons. 
\textit{First}, a global phase rotation typically has no physical consequence in most MIMO scenarios because only relative phases across ports affect interferometric functionalities such as beamforming and focusing; a global phase can be absorbed into the definitions of the reference phases. Therefore, our fidelity metric should be insensitive to a global phase rotation. 
\textit{Second}, the overall strength of $\mathbf{H}(\mathbf{r})$ depends on propagation losses within the system, and the aggregate power gain of a chosen target operator $\mathbf{H}_{\mathrm{des}}$ may be physically unattainable. In such cases, metrics based on absolute mismatch (e.g., $\left\|\mathbf{H}(\mathbf{r})-\mathbf{H}_{\mathrm{des}}\right\|_F$) conflate \emph{scale} and \emph{shape}: when $\left\|\mathbf{H}_{\mathrm{des}}\right\|_F$ greatly exceeds what is realizable (e.g., compared to $B^{\mathrm{Fro}}_{\mathrm{SDR}}$ derived in Sec.~\ref{sec_FrobeniusBounds}), the optimization is largely driven by unavoidable amplitude discrepancies rather than by the relative coupling pattern that defines the intended operator. By normalizing by $\left\|\mathbf{H}(\mathbf{r})\right\|_F^2$ in \eqref{eq:fidelity_def}, our fidelity metric becomes invariant to any nonzero complex scalar $a$ and thus directly quantifies the best attainable \emph{directional alignment} between $\mathbf{H}(\mathbf{r})$ and $\mathbf{H}_{\mathrm{des}}$ (i.e., to what extent $\mathbf{H}(\mathbf{r})$ is proportional to $\mathbf{H}_{\mathrm{des}}$), which is the notion of fidelity in operator synthesis that we seek to bound.
Concrete examples in which our fidelity metric for operator synthesis is relevant include the implementation of a required low-to-high-dimensional transformation for hybrid analog-digital beamforming with a real-life reconfigurable wave system, as well as the synthesis of radiation patterns that are principal scene components for computational meta-imaging~\cite{liang2015reconfigurable,li2019machine,saigre2022intelligent}.

Formally, the fidelity maximization problem considered in this section is summarized as follows:
\begin{equation}
\begin{aligned}
\max_{\mathbf r\in\mathbb C^{N_\mathrm S}}\quad 
& F\!\left(\mathbf{H}(\mathbf{r}),\mathbf{H}_{\mathrm{des}}\right)\\
\text{s.t.}\quad 
& r_i\in\{\alpha,\beta\},\quad i=1,\dots,N_\mathrm S.
\end{aligned}
\label{eq:fidelity_opt}
\end{equation}
The functional dependence of $F$ on $\mathbf{H}(\mathbf{r})$ and $\mathbf{H}_\mathrm{des}$ is given in (\ref{eq:fidelity_def}), and the functional dependence of $\mathbf{H}$ on $\mathbf{r}$ is given in (\ref{eq_MNT}).

Compared to the Frobenius-norm maximization problem in (\ref{eq:frob_opt}), the main difference lies in our objective. While we can reuse the constraint formulations in \eqref{eq_frob_binary_constraint} and \eqref{eq_rep_qcqp}, the fidelity objective is not a single quadratic form in $\mathbf{y}$. As seen in (\ref{eq:fidelity_def}), our objective (the fidelity metric) is the ratio of two functions that are each quadratic in $\mathbf{y}$. Therefore, we cannot formulate (\ref{eq:fidelity_opt}) as a standard QCQP, and consequently we cannot bound the achievable fidelity by directly applying the standard QCQP-to-SDP lifting from Sec.~\ref{sec_FrobeniusBounds} to the objective.

To proceed despite this difficulty, we begin by separately expressing the denominator and the numerator of the fidelity metric as quadratic forms in $\mathbf{y}$. The denominator being proportional to the Frobenius norm of $\mathbf{H}(\mathbf{r})$, we can directly reuse (\ref{eq:frob_quad_obj}):
\begin{equation}
\begin{aligned}
f_\mathrm{d}(\mathbf{y}) &= \left\|\mathbf{H}_{\mathrm{des}}\right\|_{F}^{2}\,\left\|\mathbf{H}(\mathbf{r})\right\|_{F}^{2}\\
&= h \,\mathbf y^{\dagger}\,\mathbf R_0\,\mathbf y
+2\,h\,\Re\!\left\{\mathbf q_0^{\dagger}\,\mathbf y\right\}
+h\,\tau_0,
\end{aligned}
\label{eq:fidelity_denominator_quad_obj}
\end{equation}
where $h\triangleq\left\|\mathbf{H}_{\mathrm{des}}\right\|_{F}^{2}$.

To express the numerator of \eqref{eq:fidelity_def} as a quadratic form in $\mathbf y$, we note (using two matrix identities\footnote{We use the following two matrix identities~\cite{PetersenPedersenCookbook}:\begin{enumerate}[label=(\arabic*), leftmargin=*, itemsep=0pt, topsep=2pt]
\item $\mathrm{tr}\!\left(\mathbf \Pi^{\dagger}\mathbf \Sigma\right)=\mathrm{vec}(\mathbf \Pi)^{\dagger}\,\mathrm{vec}(\mathbf \Sigma)$.
\item $\mathrm{vec}\!\left(\mathbf \Pi\,\mathbf \Sigma\,\mathbf \Lambda\right)=\left(\mathbf \Lambda^{\top}\!\otimes\!\mathbf \Pi\right)\mathrm{vec}(\mathbf \Sigma)$.
\end{enumerate}}) that
\begin{equation}
\begin{aligned}
\mathrm{tr}\!\left(\mathbf H_{\mathrm{des}}^\dagger \,\mathbf H(\mathbf r)\right)
&= c_0
+\mathrm{tr}\!\left(\mathbf H_{\mathrm{des}}^\dagger\, \mathbf A\,\mathbf X\right)
\\ &= c_0
+\mathbf q_{\mathrm{des}}^\dagger \, \mathbf y,
\end{aligned}
\end{equation}
where $c_0\triangleq \mathrm{tr}\!\left(\mathbf H_{\mathrm{des}}^\dagger \,\mathbf H_0\right)\in\mathbb C$ and $\mathbf q_{\mathrm{des}}\triangleq \mathrm{vec}\!\left(\mathbf A^\dagger\, \mathbf H_{\mathrm{des}}\right)\in\mathbb C^{N_\mathrm{S}N_\mathrm{T}}$. Upon expanding, we can now rewrite the numerator as quadratic form in $\mathbf{y}$:
\begin{equation}
\begin{aligned}
f_\mathrm{n}(\mathbf y)
&\triangleq
\left|\mathrm{tr}\!\left(\mathbf H_{\mathrm{des}}^\dagger \mathbf H(\mathbf r)\right)\right|^2\\
&=\left|c_0+\mathbf q_{\mathrm{des}}^\dagger \mathbf y\right|^2\\
&=
\mathbf y^\dagger \mathbf R_1\,\mathbf y
+2\,\Re\!\left\{\mathbf q_1^\dagger \mathbf y\right\}
+\tau_1,
\end{aligned}
\label{eq:fidelity_numerator_quad_obj}
\end{equation}
with $\mathbf R_1 \triangleq \mathbf q_{\mathrm{des}}\,\mathbf q_{\mathrm{des}}^\dagger$, $\mathbf q_1 \triangleq c_0\,\mathbf q_{\mathrm{des}}$, and $\tau_1 \triangleq |c_0|^2$.

Our fidelity-maximization objective can thus be expressed in a fractional-quadratic form in $\mathbf y$. By applying a Charnes--Cooper transformation~\cite{charnes1962programming,boyd2004convex},\footnote{The classical alternatives to the Charnes--Cooper transformation, namely (i) a bisection on a target fidelity threshold, and, (ii) Dinkelbach's method~\cite{dinkelbach1967nonlinear}, are computationally more intensive because they require multiple SDP solves.} we can rescale the variables to obtain an equivalent lifted formulation (prior to SDR) in which the scaled denominator is forced to unity. The fractional objective then reduces to the scaled numerator. After relaxing the rank-one constraint, we obtain a single SDP whose only conceptual difference relative to Sec.~\ref{sec_FrobeniusBounds} are (i) the introduction of an additional scalar scaling variable constrained to be non-negative and real-valued, and (ii) a constraint enforcing the denominator normalization.

To start, using the lifting $\mathbf Y=\mathbf y\mathbf y^\dagger$ from Sec.~\ref{sec_FrobeniusBounds}, we write the denominator \eqref{eq:fidelity_denominator_quad_obj} as
\begin{equation}
f_{\mathrm d}(\mathbf y)
= h\,\mathrm{tr}(\mathbf R_0\mathbf Y)+2\,h\,\Re\!\left\{\mathbf q_0^\dagger \mathbf y\right\}+h\,\tau_0,
\label{eq:fid_den_lifted}
\end{equation}
and the numerator \eqref{eq:fidelity_numerator_quad_obj} as
\begin{equation}
f_{\mathrm n}(\mathbf y)
= \mathrm{tr}(\mathbf R_1\mathbf Y)+2\,\Re\!\left\{\mathbf q_1^\dagger \mathbf y\right\}+\tau_1.
\label{eq:fid_num_lifted}
\end{equation}
After this lifting, our optimization problem can be expressed as
\begin{equation}
\small
\begin{aligned}
\max_{\mathbf y,\mathbf Y}\quad
& \frac{
\mathrm{tr}(\mathbf R_1\mathbf Y)
+2\,\Re\!\left\{\mathbf q_1^\dagger \mathbf y\right\}
+\tau_1
}{
h\,\mathrm{tr}(\mathbf R_0\mathbf Y)
+2\,h\,\Re\!\left\{\mathbf q_0^\dagger \mathbf y\right\}
+h\,\tau_0
}\\
\text{s.t.}\quad
& \mathrm{tr}(\mathbf R_{s,t}\mathbf Y)
+\mathbf y^\dagger \mathbf p_{s,t}
+\mathbf q_{s,t}^\dagger \mathbf y
+\tau_{s,t}=0,\\
& \hspace{1.7em}\substack{
\forall\, s\in\{1,\dots,N_\mathrm S\},\\
\forall\, t\in\{1,\dots,N_\mathrm T\},
}\\[0.2em]
& \mathrm{tr}(\mathbf R^{(i)}_{s,t}\mathbf Y)
+\mathbf y^\dagger \mathbf p^{(i)}_{s,t}
+\bigl(\mathbf q^{(i)}_{s,t}\bigr)^\dagger \mathbf y
+\tau^{(i)}_{s,t}=0,\\
& \hspace{1.7em}\substack{
\forall\, s\in\{1,\dots,N_\mathrm S\},\\
\forall\, t\in\{2,\dots,N_\mathrm T\},\\
\forall\, i\in\{1,2\},
}\\[0.2em]
& \mathbf Y=\mathbf y\mathbf y^\dagger.\\
\end{aligned}
\label{eq:fid_lifted_fractional_rank1}
\end{equation}
\normalsize
We assume that $f_\mathrm{d}(\mathbf{y})>0$ which is typically the case in practice because, as seen in~(\ref{eq:fidelity_denominator_quad_obj}), $f_\mathrm{d}(\mathbf{y})>0$ is the product of two Frobenius norms that we assume to be non-zero.

Now, we introduce the scaling variable
\begin{equation}
\sigma \triangleq \frac{1}{f_{\mathrm d}(\mathbf y)} \in \mathbb R_{>0},
\label{eq:s_def_cc}
\end{equation}
and the scaled variables
\begin{equation}
\tilde{\mathbf y}\triangleq \sigma\,\mathbf y,\qquad
\tilde{\mathbf Y}\triangleq \sigma\,\mathbf Y.
\label{eq:cc_scaled_vars}
\end{equation}
Multiplying every line in (\ref{eq:fid_lifted_fractional_rank1}) with $\sigma$, using (\ref{eq:cc_scaled_vars}), and recognizing that the objective's denominator becomes unity, we obtain
\begin{equation}
\small
\begin{aligned}
\max_{\tilde{\mathbf y},\tilde{\mathbf Y},\sigma}\quad
& \mathrm{tr}(\mathbf R_1\tilde{\mathbf Y})
+2\,\Re\!\left\{\mathbf q_1^\dagger \tilde{\mathbf y}\right\}
+\tau_1\,\sigma\\
\text{s.t.}\quad
& h\,\mathrm{tr}(\mathbf R_0\tilde{\mathbf Y})
+2h\,\Re\!\left\{\mathbf q_0^\dagger \tilde{\mathbf y}\right\}
+h\,\tau_0\,\sigma
=1,\\
& \mathrm{tr}(\mathbf R_{s,t}\tilde{\mathbf Y})
+\tilde{\mathbf y}^\dagger \mathbf p_{s,t}
+\mathbf q_{s,t}^\dagger \tilde{\mathbf y}
+\tau_{s,t}\,\sigma=0,\\
& \hspace{1.7em}\substack{
\forall\, s\in\{1,\dots,N_\mathrm S\},\\
\forall\, t\in\{1,\dots,N_\mathrm T\},
}\\[0.2em]
& \mathrm{tr}(\mathbf R^{(i)}_{s,t}\tilde{\mathbf Y})
+\tilde{\mathbf y}^\dagger \mathbf p^{(i)}_{s,t}
+\bigl(\mathbf q^{(i)}_{s,t}\bigr)^\dagger \tilde{\mathbf y}
+\tau^{(i)}_{s,t}\,\sigma=0,\\
& \hspace{1.7em}\substack{
\forall\, s\in\{1,\dots,N_\mathrm S\},\\
\forall\, t\in\{2,\dots,N_\mathrm T\},\\
\forall\, i\in\{1,2\},
}\\[0.2em]
& \tilde{\mathbf Y}=\frac{1}{\sigma}\,\tilde{\mathbf y}\tilde{\mathbf y}^\dagger,\\ 
& \sigma>0.
\end{aligned}
\label{eq:cc_exact_rank1}
\end{equation}
\normalsize
Compared to (\ref{eq:fid_lifted_fractional_rank1}), we now work in (\ref{eq:cc_exact_rank1}) with the scaled optimization variables $\tilde{\mathbf{y}}$ and $\tilde{\mathbf{Y}}$, we have the additional optimization variable $\sigma$, and we have two additional constraints related to the normalization of the denominator and the non-negativity of $\sigma$.

At this stage, the only remaining nonconvexity in (\ref{eq:cc_exact_rank1}) is the scaled rank-one relation $\tilde{\mathbf Y}=\frac{1}{\sigma}\,\tilde{\mathbf y}\tilde{\mathbf y}^\dagger$. The SDR step relaxes this equality to the linear matrix inequality $\tilde{\mathbf Y}\ \succeq\ \frac{1}{\sigma}\,\tilde{\mathbf y}\tilde{\mathbf y}^\dagger$. With $\tilde{\mathbf M}\triangleq
\begin{bmatrix}
\tilde{\mathbf Y} & \tilde{\mathbf y}\\
\tilde{\mathbf y}^\dagger & \sigma
\end{bmatrix}$, the Schur complement identity gives the equivalence $\tilde{\mathbf M}\succeq \mathbf 0  \Longleftrightarrow \tilde{\mathbf Y}\succeq \frac{1}{\sigma}\,\tilde{\mathbf y}\tilde{\mathbf y}^\dagger$. The SDR of the QCQP in \eqref{eq:cc_exact_rank1} thus becomes the following SDP:
\begin{equation}
\small
\begin{aligned}
\max_{\tilde{\mathbf y},\tilde{\mathbf Y},\sigma}\quad
& \mathrm{tr}(\mathbf R_1\tilde{\mathbf Y})
+2\,\Re\!\left\{\mathbf q_1^\dagger \tilde{\mathbf y}\right\}
+\tau_1\,\sigma\\
\text{s.t.}\quad
& h\,\mathrm{tr}(\mathbf R_0\tilde{\mathbf Y})
+2\,h\,\Re\!\left\{\mathbf q_0^\dagger \tilde{\mathbf y}\right\}
+h\,\tau_0\,\sigma
=1,\\
& \mathrm{tr}(\mathbf R_{s,t}\tilde{\mathbf Y})
+\tilde{\mathbf y}^\dagger \mathbf p_{s,t}
+\mathbf q_{s,t}^\dagger \tilde{\mathbf y}
+\tau_{s,t}\,\sigma=0,\\
& \hspace{1.7em}\substack{
\forall\, s\in\{1,\dots,N_\mathrm S\},\\
\forall\, t\in\{1,\dots,N_\mathrm T\},
}\\[0.2em]
& \mathrm{tr}(\mathbf R^{(i)}_{s,t}\tilde{\mathbf Y})
+\tilde{\mathbf y}^\dagger \mathbf p^{(i)}_{s,t}
+\bigl(\mathbf q^{(i)}_{s,t}\bigr)^\dagger \tilde{\mathbf y}
+\tau^{(i)}_{s,t}\,\sigma=0,\\
& \hspace{1.7em}\substack{
\forall\, s\in\{1,\dots,N_\mathrm S\},\\
\forall\, t\in\{2,\dots,N_\mathrm T\},\\
\forall\, i\in\{1,2\},
}\\[0.2em]
&
\begin{bmatrix}
\tilde{\mathbf Y} & \tilde{\mathbf y}\\
\tilde{\mathbf y}^\dagger & \sigma
\end{bmatrix}\succeq \mathbf 0,\\
& \sigma > 0.
\end{aligned}
\label{eq:sdr_cc_fidelity_final}
\end{equation}
\normalsize

Because every feasible point of the exact Charnes--Cooper formulation in \eqref{eq:cc_exact_rank1} is also feasible for the relaxed SDP in \eqref{eq:sdr_cc_fidelity_final}, the optimal value of the latter upper-bounds the optimal value of the former. Our SDR-based bound for the achievable fidelity is thus the optimal objective value of \eqref{eq:sdr_cc_fidelity_final}. Given a solver output \((\check{\tilde{\mathbf Y}},\check{\tilde{\mathbf y}},\check{\sigma})\) for \eqref{eq:sdr_cc_fidelity_final}, we define
\begin{equation}
B^\mathrm{Fid}_{\mathrm{SDR}}
\triangleq
\mathrm{tr}(\mathbf R_1\,\check{\tilde{\mathbf Y}})
+2\,\Re\!\left\{\mathbf q_1^{\dagger}\,\check{\tilde{\mathbf y}}\right\}
+\tau_1\,\check{\sigma}.
\label{eq:Bfid_sdr_def}
\end{equation}
The argument in the Appendix of~\cite{salmi2026electromagnetically} for the invariance of SDR-based bounds with respect to parameter ambiguities extends directly to the present case and is therefore not repeated here.

\begin{figure*}[t]
    \centering
    \includegraphics[width=2\columnwidth]{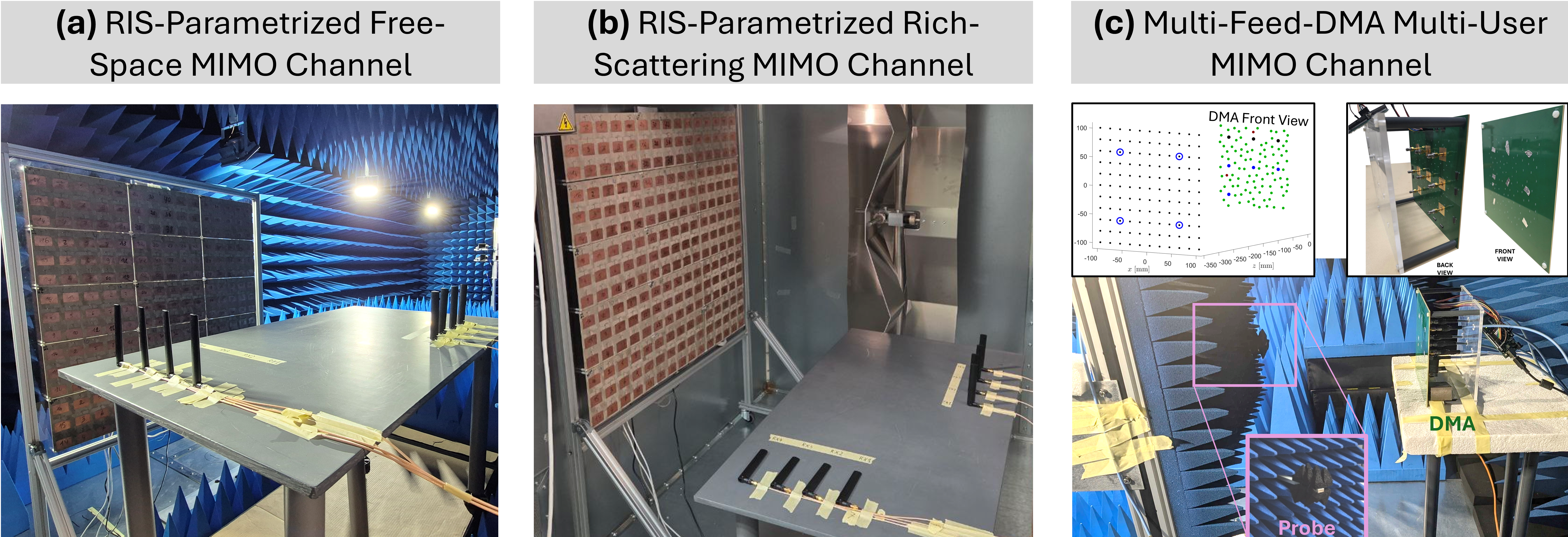}
    \caption{Experimental setups. (a) RIS-parametrized free-space MIMO channel. (b) RIS-parametrized rich-scattering MIMO channel. (c) Multi-feed-DMA multi-user MIMO channel. The four selected feeds are marked with blue dots, and the four selected user positions are highlighted with blue circles in the left inset. }
    \label{Fig1}
\end{figure*}

\section{Discrete RIS Optimization Algorithms \\to Probe the Bound Tightness}
\label{sec_Opti}

To probe the tightness of the bounds derived in Sec.~\ref{sec_FrobeniusBounds} and Sec.~\ref{sec_FidelityBounds}, we use three standard techniques for model-based discrete optimization that we briefly summarize in this section. In all cases, we begin by defining an objective function $\mathcal{O}$ that we seek to maximize. For the Frobenius-norm maximization, $\mathcal{O} = \|\mathbf H(\mathbf r)\|_F^2$, while for the fidelity maximization, $\mathcal{O} = F\!\left(\mathbf{H}(\mathbf{r}),\mathbf{H}_{\mathrm{des}}\right)$.

For an \textit{Exhaustive Search (ES)}, we evaluate $\mathcal{O}$ for each possible configuration to identify the globally best configuration. To alleviate the computational complexity, we leverage the Woodbury identity~\cite{prod2023efficient}. Nonetheless, we only use ES for $N_\mathrm{S}\leq 20$ due to its prohibitive computational cost for larger values of $N_\mathrm{S}$.

For a $\textit{Coordinate Descent (CD)}$, we initially evaluate $\mathcal{O}$ for 100 random configurations. We retain the best one as initialization and then check element by element whether flipping its state improves $\mathcal{O}$. We loop multiple times over all elements, until convergence when the last $N_\mathrm{S}$ iterations did not improve $\mathcal{O}$. We alleviate the computational cost again using the Woodbury identity~\cite{prod2023efficient}.

For the \textit{Genetic Algorithm (GA)}, we use $\mathcal{O}$ as the fitness function.  We use default parameters (population size: 200, maximum number of generations: $100 N_\mathrm{S}$) and stop early if there is no improvement for the last 50 generations or if the improvement drops below a threshold of $10^{-6}$.

In addition to these three standard techniques for discrete optimization, we attempt to extract an optimized configuration from the solution of the SDP underlying the SDR-based bound. To find the \textit{Projected SDR Solution (P-SDR)}, we obtain a feasible configuration from the (usually non-rank-one) optimizer $(\check{\mathbf Y},\check{\mathbf y})$ (i.e., $\check{\mathbf Y}\neq \check{\mathbf y}\,\check{\mathbf y}^\dagger$ in general). Specifically, we first reshape the relaxed vector solution into a matrix $\check{\mathbf X}=\mathrm{unvec}(\check{\mathbf y})\in\mathbb C^{N_\mathrm S\times N_\mathrm T}$. Then, we compute $\check{\mathbf Z}=\mathbf B+\mathbf\Gamma\check{\mathbf X}$. Finally, we choose the binary state of each tunable element $s\in\{1,\dots,N_\mathrm S\}$ by comparing which state ($\alpha$ or $\beta$) better matches the relaxed solution across all transmit indices $t=1,\dots,N_\mathrm T$. Concretely, we define the row-wise residuals $R_\alpha(s)\triangleq\|\mathbf u_s^{\top}\check{\mathbf X}-\alpha\,\mathbf u_s^{\top}\check{\mathbf Z}\|_2$ and $R_\beta(s)\triangleq\|\mathbf u_s^{\top}\check{\mathbf X}-\beta\,\mathbf u_s^{\top}\check{\mathbf Z}\|_2$, where $\mathbf u_s^{\top}\check{\mathbf X}\in\mathbb C^{1\times N_\mathrm T}$ and $\mathbf u_s^{\top}\check{\mathbf Z}\in\mathbb C^{1\times N_\mathrm T}$ extract the $s$th rows of $\check{\mathbf X}$ and $\check{\mathbf Z}$, respectively. We obtain a binary control vector $\check{\mathbf v} \in \mathbb{B}^{N_\mathrm{S}}$ by setting $\check v_s=1$ if $R_\beta(s)<R_\alpha(s)$ and $\check v_s=0$ otherwise.

\section{Experimental Results}
\label{sec_Experimental}

In this section, we evaluate our Frobenius-norm bound from Sec.~\ref{sec_FrobeniusBounds} and our fidelity bound from Sec.~\ref{sec_FidelityBounds} for three real-world experimental systems. We describe the considered experimental setups in Sec.~\ref{subsec_ExpSetups}. Then, we evaluate our bounds and probe their tightness with the discrete optimization techniques summarized in Sec.~\ref{sec_Opti}.

\subsection{Experimental Setups}
\label{subsec_ExpSetups}

In the following, we describe our three distinct experimental setups. The first two setups are both RIS-parametrized wireless channels based on the same RIS prototype, but the environmental scattering drastically differs. Previous work has shown that these differences in environmental scattering substantially influence the strength of the mutual coupling  between the RIS elements~\cite{rabault2023tacit,del2025experimentalreducedrank}. 

\textit{(a) RIS-Parametrized Free-Space MIMO Channel:}
We consider a $4 \times 4 $ MIMO channel inside an anechoic chamber that is parametrized by 100 1-bit-programmable RIS elements, as displayed in Fig.~\ref{Fig1}a. The RIS prototype consists of 225 identical elements, of which we use only 100; the remaining 115 elements are fixed to a reference configuration throughout all experiments. The RIS elements are designed for operation at 2.45~GHz. Each RIS element is equipped with a PIN diode that endows it with 1-bit-programmability. The validity of the MNT description, which describes each electrically very small PIN diode as a lumped element, has been verified in prior work~\cite{largeRIS_TCOM,del2025frozen}. Further details on the RIS design can be found in~\cite{ahmed2025over}. The transmit array and the receive array are identical and comprise four commercial WiFi antennas (ANT-W63WS2-SMA) with regular half-wavelength spacing. In this setup, both antenna arrays' orientations are aligned with that of the RIS elements. We emphasize that our system model makes no assumption about the antennas' orientations and locations.
Altogether, $N_\mathrm{T}=N_\mathrm{R}=4$ and $N_\mathrm{S}=100$ for this system; we have estimated the system model parameters at 2.45~GHz using the technique described in~\cite{ContRIS_LWC}.

\textit{(b) RIS-Parametrized Rich-Scattering MIMO Channel:}
We consider a $4 \times 4 $ MIMO channel inside a reverberation chamber that is parametrized by 100 1-bit-programmable RIS elements, as displayed in Fig.~\ref{Fig1}b.  The RIS prototype and the antenna arrays are the same as in (a). The radio environment features rich environmental scattering, being a reverberation chamber (1.75~$\times$~1.5~$\times$~2~$\mathrm{m}^3$). The chamber's mode stirrer is kept in a fixed position throughout all experiments. We emphasize that our system model makes no assumptions about the antennas' orientations or locations, nor about that the complexity of the environmental scattering. In this setup, the orientation of the transmit array is aligned with the RIS elements' orientation, while the receive array's orientation is perpendicular to that of the transmit array to minimize the direct link. As already mentioned, our system model makes no assumption about the antennas' orientations and locations.  Altogether, $N_\mathrm{T}=N_\mathrm{R}=4$ and $N_\mathrm{S}=100$ for this system; we have estimated the system model parameters at 2.45~GHz using the technique described in~\cite{ContRIS_LWC}.

\textit{(c) Multi-Feed-DMA Multi-User MIMO Channel:}
We consider a $4 \times 4 $ MIMO channel inside an anechoic chamber, from four feeds of a DMA to four users, as displayed in Fig.~\ref{Fig1}c; the channel is parametrized by 96 meta-elements within the DMA. Specifically, the DMA comprises eight feeds and 96 1-bit-programmable meta-elements that are coupled via a quasi-2D chaotic cavity. Four feeds are unused in our experiments (one feed is assumed to be terminated in an open-circuit load, and the three other feeds are assumed to be terminated in matched loads). The DMA architecture follows the design described in~\cite{sleasman2020implementation} except that it has multiple feeds instead of a single feed, as described in~\cite{tapie2025experimental}. The meta-elements are designed for operation in the lower K-band following~\cite{yoo2016efficient}; each meta-element is based on a complementary electric-LC (cELC) resonator equipped with a PIN diode that endows it with 1-bit-programmability. The validity of the MNT description, which describes each electrically very small PIN diode as a lumped element, has been verified in prior work~\cite{tapie2025experimental}. 
A 2D translation stage displaces an open-ended waveguide probe (42EWGS-A1, A-INFO INC.) in a plane parallel to the DMA at a distance of 35~cm from the DMA's front. The four selected user positions are marked with blue circles in Fig.~\ref{Fig1}c.
Further details on the utilized DMA design and experimental setup can be found in~\cite{tapie2025experimental}. Altogether, $N_\mathrm{T}=N_\mathrm{R}=4$ and $N_\mathrm{S}=96$ for this system; we have estimated the system model parameters at 19~GHz using the technique described in~\cite{tapie2025experimental}.

When we work with a value of $N_\mathrm{S}$ below 100 for (a) and (b) or below 96 for (c), we fix the termination of the remaining unused tunable elements to $\alpha$.\footnote{We split the set $\mathcal S$ of ``virtual'' ports into $\mathcal S_1$ ($N_{\mathrm S}'$ used elements) and
$\mathcal S_2$ ($N_{\mathrm S}-N_{\mathrm S}'$ unused elements) and partition the tunable-related blocks accordingly as
$\mathbf A=\bigl[\mathbf A_{\mathcal S_1}\ \mathbf A_{\mathcal S_2}\bigr]$,
$\mathbf B=\begin{bmatrix}\mathbf B_{\mathcal S_1}\\ \mathbf B_{\mathcal S_2}\end{bmatrix}$,
$\mathbf\Gamma=\begin{bmatrix}\mathbf\Gamma_{\mathcal S_1\mathcal S_1}&\mathbf\Gamma_{\mathcal S_1\mathcal S_2}\\
\mathbf\Gamma_{\mathcal S_2\mathcal S_1}&\mathbf\Gamma_{\mathcal S_2\mathcal S_2}\end{bmatrix}$.
Fixing the unused loads to $\alpha$ corresponds to $\mathbf\Phi_{\mathcal S_2}=\alpha\,\mathbf I_{\tilde{n}}$, where $\tilde{n} = N_\mathrm{S} - N_\mathrm{S}^\prime$, and yields a reduced MNT model of the form \eqref{eq_MNT} with  
\begin{align*}
\mathbf\Gamma^\prime \;&=\;\mathbf\Gamma_{\mathcal S_1\mathcal S_1}+\mathbf\Gamma_{\mathcal S_1\mathcal S_2}(\mathbf I_{\tilde{n}}-\mathbf\Phi_{\mathcal S_2}\mathbf\Gamma_{\mathcal S_2\mathcal S_2})^{-1}\mathbf\Phi_{\mathcal S_2}\mathbf\Gamma_{\mathcal S_2\mathcal S_1},\\
\mathbf B^\prime \;&=\;\mathbf B_{\mathcal S_1}+\mathbf\Gamma_{\mathcal S_1\mathcal S_2}(\mathbf I_{\tilde{n}}-\mathbf\Phi_{\mathcal S_2}\mathbf\Gamma_{\mathcal S_2\mathcal S_2})^{-1}\mathbf\Phi_{\mathcal S_2}\mathbf B_{\mathcal S_2},\\
\mathbf A^\prime \;&=\;\mathbf A_{\mathcal S_1}+\mathbf A_{\mathcal S_2}(\mathbf I_{\tilde{n}}-\mathbf\Phi_{\mathcal S_2}\mathbf\Gamma_{\mathcal S_2\mathcal S_2})^{-1}\mathbf\Phi_{\mathcal S_2}\mathbf\Gamma_{\mathcal S_2\mathcal S_1},\\
\mathbf H_0^\prime \;&=\;\mathbf H_0+\mathbf A_{\mathcal S_2}(\mathbf I_{\tilde{n}}-\mathbf\Phi_{\mathcal S_2}\mathbf\Gamma_{\mathcal S_2\mathcal S_2})^{-1}\mathbf\Phi_{\mathcal S_2}\mathbf B_{\mathcal S_2}.
\end{align*}
}

Comparing the three setups in terms of the inter-element coupling strength is challenging because metrics directly based on $\mathbf{S}_\mathcal{SS}$ are sensitive to parameter ambiguities. Indeed, such metrics can be altered with a gauge transformation of the parameters that leaves the $\mathbf{v}\rightarrow \mathbf{H}(\mathbf{v})$ mapping unchanged. Even if the system was perfectly known, redefining the reference impedance at the ports associated with the tunable elements (which is closely related to the diagonal-similarity gauge described in~\cite{salmi2026electromagnetically}) would change such metrics. Thus, a metric evaluated based on the unambiguous end-to-end channel is preferable. One such metric examines the accuracy with which a simple multi-variable linear regression can map $\mathbf{v}$ to $\mathbf{H}$~\cite{rabault2023tacit}. In the absence of inter-element coupling (and assuming 1-bit-programmable elements), the true mapping from $\mathbf{v}$ to $\mathbf{H}$ is affine. Based on prior works on our experimental systems~\cite{rabault2023tacit,del2025experimentalreducedrank,tapie2025experimental}, the inter-element coupling is weak in (a), moderately strong in (b), and very strong in (c).

\begin{figure*}
    \centering
    \includegraphics[width=2\columnwidth]{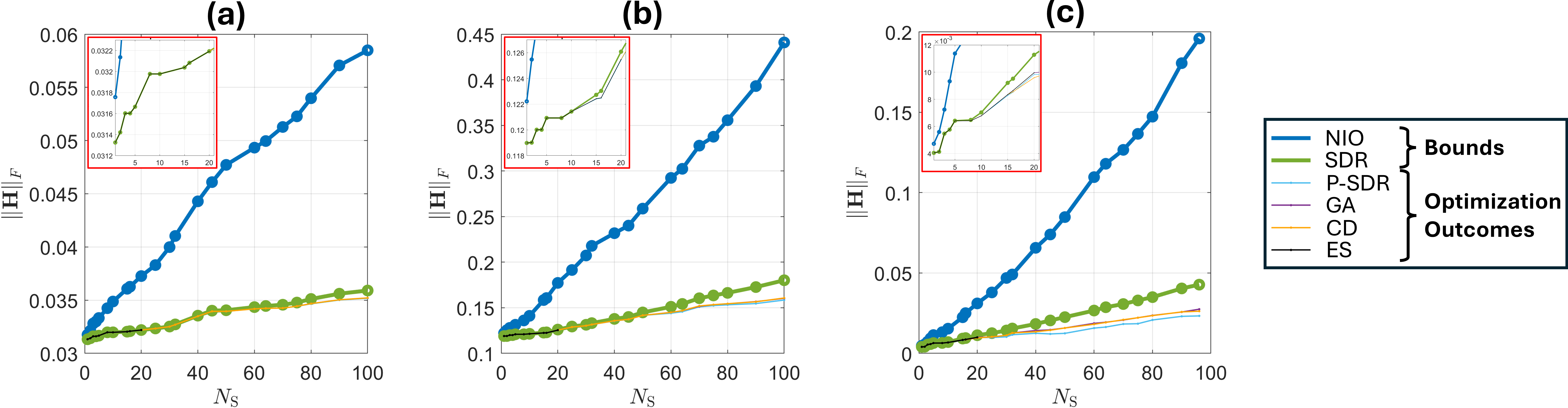}
    \caption{NIO (blue) and SDR (green) bounds on the achievable Frobenius norm of the end-to-end channel matrix as a function of $N_\mathrm{S}$, for the three experimental setups in Fig.~\ref{Fig1}. The bounds are contrasted with the outcomes of four discrete optimizations (thin lines - see Sec.~\ref{sec_Opti}). The insets show a zoom on small values of $N_\mathrm{S}$. }
    \label{Fig2}
\end{figure*}

\subsection{Experimental Results: Frobenius-Norm Maximization}
\label{subsec_ExpResults_FrobMax}

Our evaluation of the NIO and SDR bounds on the largest achievable Frobenius norm (derived in Sec.~\ref{sec_FrobeniusBounds}) for our three experimental setups is displayed in Fig.~\ref{Fig2} as a function of $N_\mathrm{S}$, and contrasted with the outcomes of the four discrete optimization techniques described in Sec.~\ref{sec_Opti}. Upon visual inspection, it is clear that the SDR bound is notably tighter than the NIO bound in all cases. In fact, our discrete optimizations almost perfectly reach the SDR bound up to $N_\mathrm{S}\approx 45$ in (a), up to $N_\mathrm{S}\approx 25$ in (b), and up to $N_\mathrm{S}\approx 8$ in (c). Even at the largest considered values of $N_\mathrm{S}$, our discrete optimizations reach 96\ \% of our SDR bound in (a), 79\ \% of our SDR bound in (b), and 41\ \% of our SDR bound in (c). The ability of the discrete optimization techniques to closely approach the SDR bound thus appears to depend on $N_\mathrm{S}$ and on the strength of inter-element coupling. We emphasize that observing a gap between a bound and a discrete-optimization outcome is not clear evidence for a loose bound, since this observation alone does not rule out that the discrete optimization was inefficient and unable to approach the global optimum. In fact, we do expect that discrete optimization becomes more challenging for larger $N_\mathrm{S}$ and systems with stronger inter-element coupling. Only for values of $N_\mathrm{S}$ up to 20 we know the global optimum from ES. As seen in the inset, for $N_\mathrm{S}=20$, the SDR bound perfectly coincides with the global optimum in (a), the SDR bound exceeds the global optimum by 0.5\ \% in (b), and the SDR bound exceeds the global optimum by 13.3\ \% in (c). The SDR bound's tightness thus displays a significant dependence on the inter-element coupling strength. For the case of very strong inter-element coupling in (c),  we can conclude that the SDR bound is not perfectly tight for $N_\mathrm{S}=20$ because ES identifies the global optimum. 

We summarize in Table~\ref{tab:bound_tightness_exp_frob} how closely the best out of the considered discrete-optimization techniques approaches the NIO and SDR bounds, for the largest considered values of $N_\mathrm{S}$ in each setup. Altogether, it is apparent that the SDR bound yields remarkably tight bounds. In case (a) with weak inter-element coupling, our SDR bound exceeds the global optimum by \textit{at most} a factor of 1.04; based on our SDR bound, we can thus guarantee that no other algorithms capable of finding substantially better optimization outcomes can exist. Even for cases (b) and (c) with stronger inter-element coupling, the SDR bound is reasonably tight. Based on our discrete-optimization outcomes, we can state that the SDR bound exceeds the global optimum by \textit{at most} a factor of 1.26 in (b) and 2.42 in (c). Because the discrete optimizations are likely inefficient, the precise factor by which the SDR bound exceeds the global optimum is likely below these values. In any case, we can conclude that the SDR bound's tightness is significantly better than an order of magnitude in all considered experimental setups, which is a remarkable result considering the complexity of the underlying optimization problems ($2^{N_\mathrm{S}}$ possible system configurations and highly intertwined effects of the optimizable parameters on the objective function). In contrast, the comparison of the simpler NIO bound to the SDR bound reveals that the NIO bound's tightness is roughly an order of magnitude worse in (b) and roughly two orders of magnitude worse in (c). We additionally display in Table~\ref{tab:bound_tightness_exp_frob} the effective rank~\cite{roy2007effective} of the SDR optimizer for each case. As expected, this metric does not clearly predict the tightness of the SDR bound.

\begin{table}
\caption{NIO and SDR bounds relative to the best optimized Frobenius objective $\|\mathbf{H}_{\mathrm{opt}}\|_F^2$, as well as effective rank of the SDR optimizer, for the largest considered value of $N_\mathrm{S}$ in each setup.}
\label{tab:bound_tightness_exp_frob}
\centering
\renewcommand{\arraystretch}{1.15}
\begin{tabular}{cccc}
\hline
Setup & $B^\mathrm{Fro}_{\mathrm{NIO}}/\|\mathbf{H}_{\mathrm{opt}}\|_F^2$ & $B^\mathrm{Fro}_{\mathrm{SDR}}/\|\mathbf{H}_{\mathrm{opt}}\|_F^2$ & $R_{\mathrm{eff}}(\check{\mathbf Y})$\\
\hline
(a) & 2.76 & 1.04 & 1.58 \\
(b) & 7.53 & 1.26 & 7.42 \\
(c) & 50.85 & 2.42 & 5.72 \\
\hline
\end{tabular}
\end{table}

\subsection{Experimental Results: Fidelity Maximization}
\label{subsec_ExpResults_Fidelity}

\begin{figure*}
    \centering
    \includegraphics[width=2\columnwidth]{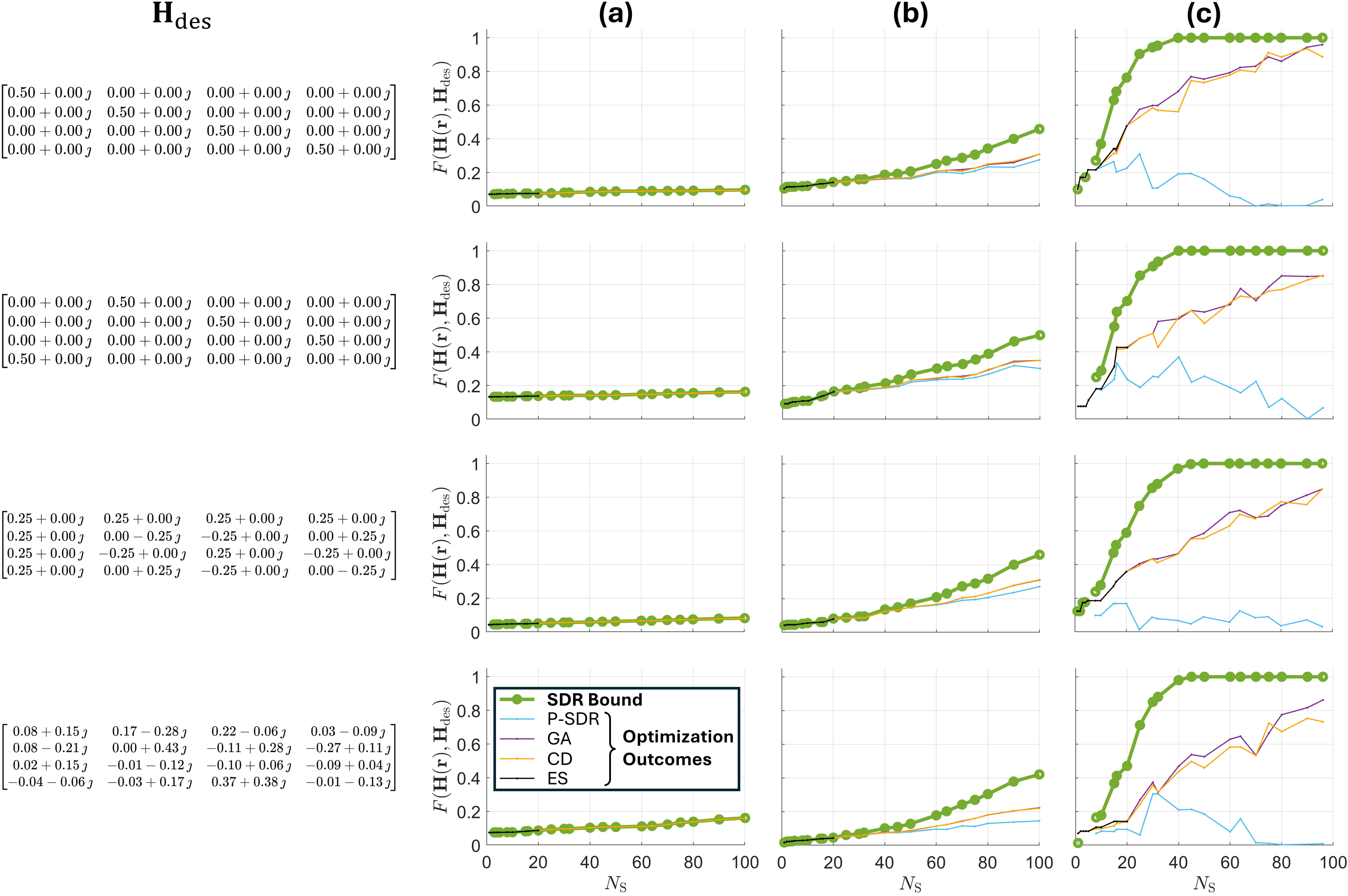}
    \caption{SDR bounds on the achievable operator-synthesis fidelity (see (\ref{eq:fidelity_def}) for definition) for four different $\mathbf{H}_\mathrm{des}$ (indicated on the left) for the three considered experimental setups shown in Fig.~\ref{Fig1}, and optimization outcomes trying to approach the bound. The four targeted $\mathbf{H}_\mathrm{des}$ operators are a scaled identity, a cyclic permutation, a scaled 4-point discrete-Fourier-transform matrix, and a random matrix. }
    \label{Fig3}
\end{figure*}

Our evaluation of the SDR bound on the achievable fidelity (derived in Sec.~\ref{sec_FidelityBounds}) for our three experimental setups is displayed in Fig.~\ref{Fig3} as a function of $N_\mathrm{S}$ for four different $\mathbf{H}_\mathrm{des}$, and contrasted with the outcomes of the four discrete optimization techniques described in Sec.~\ref{sec_Opti}. We observe substantial differences across the three setups. In (a) with weak inter-element mutual coupling, the bound appears almost perfectly tight for all values of $N_\mathrm{S}$ because our discrete optimizations reach the bound. At the same time, even with $N_\mathrm{S}=100$, the achievable fidelity remains very low for all four considered examples of $\mathbf{H}_\mathrm{des}$. In other words, our bound certifies that it is impossible to configure the system to approximate any of the four considered operators with a decent fidelity. In (b) with intermediate inter-element coupling, the bound on the fidelity is tight for low values of $N_\mathrm{S}$ (because our optimization outcomes reach it). For the largest value of $N_\mathrm{S}=100$, our bound suggests in all four cases that fidelities cannot exceed values in the vicinity of 0.45 (the exact value slightly varies with $\mathbf{H}_\mathrm{des}$), and our best optimization outcomes reach fidelities between 0.2 and 0.35 (again, the exact values vary with $\mathbf{H}_\mathrm{des}$). While we cannot distinguish between our bound being looser for large $N_\mathrm{S}$ or our discrete-optimization techniques being less efficient, the bound clarifies that it is impossible to reach a high fidelity (e.g., above 0.9) with the system in (b) for all four considered target operators. This definite conclusion is of significant practical value. In (c) with strong inter-element coupling, the bound quickly reaches unity as $N_\mathrm{S}$ approaches 40. Although unity is the trivial fidelity upper bound, our SDR bound thus does not exclude the possibility that we can reach high fidelities in (c), and, indeed, the best optimization outcomes reach fidelities up to 0.96 in (c). For a few small values of $N_\mathrm{S}$, the SDP solver fails in (c) because $\sigma$ becomes excessively large, leading to numerical instability and causing the solver to diverge. For this reason, the line for the SDR-based bound in Fig.~\ref{Fig3}c is sometimes interrupted at low values of $N_\mathrm{S}$. In these cases, the alternative approach based on bisection over a target fidelity threshold converges to the trivial bound of unity and therefore cannot serve as a reliable replacement for the Charnes–Cooper formulation.

Altogether, the SDR bound thus yields valuable insights in terms of achievable fidelities for realizing a desired operator in a given experimental system with wave-domain programmability. Incidentally, the strong dependence of both the upper fidelity bounds and the fidelity values achieved with discrete optimization confirms that strong inter-element coupling substantially boosts the control over the system's transfer function. All three considered systems have approximately the same number of 1-bit-programmable elements, but the wave-domain flexibility evidenced by the ability to implement a desired linear operator clearly varies with the inter-element coupling strength. This observation is in line with recent findings in~\cite{WCM_MC,MC_benefits_TAP} which emphasize that stronger inter-element coupling implies that the wave bounces more often between the tunable elements and thus becomes more sensitive to their configuration.

\section{Discussion}
\label{sec_discussion}

We have evaluated our bounds on the achievable aggregate channel gain and the achievable operator-synthesis fidelity in three experimental systems, spanning from weak via moderately strong to very strong inter-element coupling. We evaluated our bounds in two different frequency regimes, namely at 2.45~GHz for (a) and (b) and at 19~GHz for (c), and in two different system architectures involving programmable metasurfaces, namely a RIS-parametrized radio environment in (a) and (b) and a DMA in (c). Thereby, we have demonstrated an important feature of our bounds, namely the ability to evaluate them for diverse examples of real-life programmable-wave-domain hardware. Our underlying system model only assumes that the system is static except for 1-bit-tunable lumped elements. Extensions to multi-bit tunability are possible in future work, but existing prototypes with hundreds of tunable elements typically limit their elements to 1-bit tunability in order to alleviate control-circuit complexity.
In particular, we make no assumptions about the orientation, the location, or the structural design of the antennas, nor do we make any simplifying assumptions (e.g., unilateral approximation, ignoring mutual coupling between tunable elements, ignoring attenuation of tunable elements, ignoring the limited programmability of tunable elements). Consequently, our technique can be applied to most real-life embodiments of contemporary wave systems based on programmable metasurfaces. In particular, we note that the tunability of ``beyond-diagonal'' RISs~\cite{shen2021modeling} and ``beyond-diagonal'' DMAs typically relies on tunable lumped elements, such that our system model applies, as discussed in~\cite{del2025physics,bddma}. Indeed, a system involving beyond-diagonal RISs (and/or stacked RISs) can be partitioned into static components and tunable lumped elements~\cite{del2025physics}, and is thus perfectly compatible with our system model.

\section{Conclusion}
\label{sec_conclusion}

To summarize, we have derived fundamental bounds on the achievable aggregate channel gain and the achievable operator-synthesis fidelity in real-world reconfigurable MIMO wave systems. We applied our bounds to three distinct systems based on RISs and DMAs, and systematically studied the bounds' scaling with $N_\mathrm{S}$ and the inter-element coupling strength. 
Looking forward, the presented bounding strategy can be extended to more complex objectives, for instance, information-theoretic metrics that explicitly involve the singular values of the end-to-end MIMO transfer function.

\section*{Acknowledgment}
P.d.H. acknowledges stimulating discussions with O.~D.~Miller, S.~Molesky, A.~Salmi, and V.~Viikari. P.d.H. further acknowledges I.~Ahmed, F. Boutet, and C. Guitton as well as J.~Tapie, who, under P.d.H.'s supervision, previously built the RIS prototype for the work presented in~\cite{ahmed2025over} and the DMA prototype for the work presented in~\cite{tapie2025experimental}, respectively. Moreover, P.d.H. acknowledges J.~Sol, who provided technical support for setting up the experiments at IETR's QOSC test facility (which is part of the CNRS RF-Net network).

\bibliographystyle{IEEEtran}

\providecommand{\noopsort}[1]{}\providecommand{\singleletter}[1]{#1}%

\end{document}